\documentclass[11pt]{article}
\usepackage{amsmath,amsfonts,amssymb,latexsym}
\usepackage{hhline}
\usepackage{graphicx}
\usepackage[arrow,matrix]{xy}
\usepackage{epstopdf}

\allowdisplaybreaks[1]

\newlength{\fighskip} \fighskip=2pt
\newlength{\figvskip} \figvskip=3pt

\newcommand*{\figbox}[2]{{
  \def\figscale{#1}
  \def\arraystretch{0.8}
  \arraycolsep=0pt
  \begin{array}{c}
    \vbox{\vskip\figscale\figvskip
      \hbox{\hskip\figscale\fighskip
        \includegraphics[scale=\figscale]{#2}}}
  \end{array}}}

\newcommand\void[1]       {}

\newcommand\be            {\begin{equation}}
\newcommand\ee            {\end{equation}}
\newcommand\bea       {\begin{eqnarray}}
\newcommand\eea       {\end{eqnarray}}

\newcommand{\calB}{\mathcal{B}}
\newcommand{\calC}{\mathcal{C}}
\newcommand{\calD}{\mathcal{D}}
\newcommand{\calE}{\mathcal{E}}
\newcommand{\calF}{\mathcal{F}}
\newcommand{\calG}{\mathcal{G}}
\newcommand{\calH}{\mathcal{H}}
\newcommand{\calK}{\mathcal{K}}
\newcommand{\calL}{\mathcal{L}}
\newcommand{\calM}{\mathcal{M}}
\newcommand{\calN}{\mathcal{N}}

\newcommand{\ZZ}{\mathbb{Z}}

\newcommand{\Hilb}{\mathrm{Hilb}}
\newcommand{\Rep}{\mathop{\mathrm{Rep}}\nolimits}
\newcommand{\Hom}{\mathop{\mathrm{Hom}}\nolimits}

\newcommand{\Fun}{\mathop{\mathrm{Fun}}\nolimits}

\newcommand{\eval}{\mathop{\mathrm{eval}}\nolimits}

\newcommand{\one}{\textbf{1}}
\newcommand{\id}{\mathop{\mathrm{id}}\nolimits}

\newcommand{\dual}{\overline}

\newcommand{\M}{\mathcal{M}}
\newcommand{\op}{\mathrm{op}}
\newcommand{\opt}{\leftrightarrow}
\newcommand{\mult}{\bullet}

\newcommand*{\eqtitle}[1]{\text{\underline{#1}:}}

\title{\vskip-0.7cm Models for gapped boundaries and domain walls}
\author{Alexei Kitaev$\,{}^{a}$, Liang Kong$\,{}^{b}$,\\
  {\normalsize\it  $^{a}$ California Institute of Technology, Pasadena, CA, 91125, USA} \\
  {\normalsize\it $^{b}$ Institute for Advanced Study, Tsinghua University, Beijing, 100084, China.   
  }}

\date{}

\begin{document}
\maketitle

\begin{abstract}
We define a class of lattice models for two-dimensional topological phases with boundary such that both the bulk and the boundary excitations are gapped. The bulk part is constructed using a unitary tensor category $\calC$ as in the Levin-Wen model, whereas the boundary is associated with a module category over $\calC$. We also consider domain walls (or defect lines) between different bulk phases. A domain wall is transparent to bulk excitations if the corresponding unitary tensor categories are Morita equivalent. Defects of higher codimension will also be studied. In summary, we give a dictionary between physical ingredients of lattice models and tensor-categorical notions. 
\end{abstract}

\section{Introduction}

Relations between the bulk and the boundary have proved important for the understanding of quantum Hall states. For example, the bulk electron wave function for the Moore-Read state~\cite{moore-read} is constructed using conformal blocks of a certain conformal field theory, which also describes the edge modes (under suitable boundary conditions). Based on the success of this and similar theories, one might erroneously conclude that the bulk-boundary correspondence is one-to-one. It is, however, known that the boundary properties are generally richer than those of the bulk; in particular, the same bulk can have different boundaries. This phenomenon appears in its basic form when both the bulk and the boundary are gapped\footnote{A Hamiltonian is called ``gapped" if the smallest excitation energy, i.e.\ the difference between the two lowest eigenvalues, is bounded from below by a constant that is independent of the system size. In many cases, it is possible to define a local energy gap, which is measured between the ground state and the excited states that can be obtained by the action of local operators.}, but it should be relevant to some quantum Hall states as well.

A simple example of a topological phase that admits a gapped boundary is a $\ZZ_2$ gauge theory. Its Hamiltonian realizations include certain dimer models~\cite{MoessnerSondhi,msp}. Read and Chakraborty \cite{ReadChakraborty} studied the quasiparticle statistics and other topological properties of the $\ZZ_2$ phase. An exactly solvable Hamiltonian in this universality class (the ``toric code'' model) was proposed by the first author \cite{kitaev97}. Already in this simple example, as shown by Bravyi and Kitaev~\cite{bk98}, the bulk ``toric code'' system has two topologically distinct boundary types.

An analogue of the toric code for an arbitrary finite group $G$ was also proposed in~\cite{kitaev97}. Levin and Wen~\cite{lw-mod} went even further, replacing the group (or, rather, its representation theory) by a {\it unitary tensor category}\footnote{More exactly, a unitary finite spherical fusion category.}. Both models may be viewed as Hamiltonian realizations of certain TQFTs (or state sums in the sense of Turaev and Viro~\cite{TV}), which were originally introduced to define 3-manifold invariants. Thus, the Kitaev model corresponds to a special case of the Kuperberg invariant~\cite{kuperberg} (the general case was considered in~\cite{bmca}), whereas the Levin-Wen model corresponds to the Barrett-Westbury invariant~\cite{bw}.

Boundaries for the Kitaev model have been studied recently~\cite{bsw}. In this paper, we will outline our constructions of all possible boundaries and defects in Levin-Wen models. The details will appear in \cite{kk}. We cannot readily defend the word ``all'' in this claim since our method is limited to a particular class of models. As a parallel development, boundary conditions for Abelian Chern-Simons theories have also been characterized~\cite{kapustin-saulina-1,kapustin-saulina-2}. However, an alternative approach is possible, where one postulates some general properties (such as the fusion of quasiparticles) and studies algebraic structures defined by those axioms. This idea has long been implemented for bulk 2d systems~\cite{FRS,FG}, with the conclusion that the quasiparticles are characterized by a \emph{unitary modular category} (see Appendix~E in Ref.~\cite{kitaev06} for review). A similar theory of gapped boundaries has been contemplated by the first author and will appear in a separate paper.

In the Levin-Wen model associated with a unitary tensor category $\calC$, the bulk excitations are objects of the unitary modular category $Z(\calC)$, the monoidal \emph{center} of $\calC$ (a generalization of Drinfeld's double). This result follows from the original analysis by Levin and Wen, but we will derive it from a theory of excitations on a domain wall between two phases. Indeed, bulk excitations may be viewed as excitations on a trivial domain wall between two regions of the same phase. In the simpler case of a standard boundary between the Levin-Wen model and vacuum, the excitations are objects of the category $\calC$. Thus, the boundary theory uniquely determines the bulk theory by taking the monoidal center. On the other hand, the bulk can not completely determine the boundary because the same modular category may be realized as the center of different tensor categories, say, $\calC$ and $\calD$. Nevertheless, the bulk theory uniquely determines the boundary theory up to Morita equivalence. This is the full content of the bulk-boundary duality in the framework of Levin-Wen models. We will explicitly construct a $\calD$ boundary for the $\calC$ Levin-Wen model using the notion of a module over a tensor category.

Besides bulk-boundary duality, we also emphasize an interesting correspondence between the dualities among bulk theories (as braided monoidal equivalences) and ``transparent'', or ``invertible'' domain walls (or defect lines). In particular, for Morita equivalent $\calC$ and $\calD$, we will construct a transparent domain wall between the $\calC$ and $\calD$ models.  One can see explicitly how excitations in one region tunnel through the wall into the other region, which is just another lattice realization of the same phase. In the mathematical language, this tunneling process gives a braided monoidal equivalence between $Z(\calC)$ and $Z(\calD)$.  Moreover, the correspondence between transparent domain walls and equivalences of bulk theories is bijective\footnote{It turns out that this exact correspondence is not an isolated phenomenon. For example, a similar result holds for rational conformal field theory \cite{dkr}. We believe that it is a manifestation of a more general principle shared by many quantum field theories.} (see Section \ref{sec:lw-defect} for more precise statement). If $\calC$ and $\calD$ are themselves equivalent (as monoidal categories), the domain wall can terminate, and the transport of excitations around the endpoint defines an automorphism of $Z(\calC)$. The possibility of quasiparticles changing their type due to a transport around a point-like defect was mentioned in~\cite{kitaev06}. Such defects were explicitly constructed and studied by Bombin~\cite{bombin} under the name of ``twists'', though in his interpretation the associated domain wall is immaterial (like a Dirac string). General domain walls between phases are not transparent. A particle injected into a foreign phase leaves behind a trace (a superposition of domain walls). This process is also described using tensor category theory.

Another result of our work is a uniform treatment of different excitation types. As already mentioned, bulk quasiparticles are equivalent to excitations on the trivial domain wall. A wall between two models $\calC$ and $\calD$ can be regarded as a boundary of a single phase $\calC\boxtimes\calD^\opt$ if we fold the plane. Thus, it is sufficient to consider boundary excitations. We characterize them as superselection sectors (or irreducible modules) of a local operator algebra (see Section \ref{sec:lw-boundary}). This construction provides a crucial link between the physically motivated notion of excitation and more abstract mathematical concepts. We will also show how the properties of boundary excitations can be translated into tensor-categorical language, which leads to the mathematical notion of a module functor. This view of excitations also works perfectly well for boundary points between different types of domain walls. They can be described by more general module functors. A domain wall (also called a defect line) has codimension 1; an excitation connecting multiple domain walls is a defect of codimension 2. One can go further to consider defects of codimension 3, which are given by natural transformations between module functors. A Levin-Wen model together with defects of codimension 1, 2, 3 provides the physical meaning behinds the so-called extended Turaev-Viro topological field theory \cite{TV, lurie, kirillov-balsam, kapustin}.\smallskip

The representation theory of unitary tensor categories is central to our construction. Unfortunately, it is too rich to be covered in this short paper. Instead of giving detailed definitions of the mathematical notions that are involved, we will only motivate and explain the basic ingredients of some crucial concepts such as a module over a unitary tensor category and a module functor. We will also give a dictionary between physical ingredients in Levin-Wen models and tensor-categorical notions in Section \ref{sec:dictionary}.\smallskip

The outline of this paper is as follows:  in Section 2, we describe gapped boundaries and a domain wall in a concrete model, the toric code; in Section 3, we briefly review the Levin-Wen model; in Section 4, we outline the construction of a gapped boundary and study boundary excitations; in Section 5, we study domain walls and the tunneling of bulk excitations through the wall; in Section 6, we discuss defects of higher codimensions. In Section 7, we summarize the results by giving a translation between physical terminology (in the context of Levin-Wen models) and tensor-categorical notions; we also discuss some possible generalizations. 

\bigskip
\noindent {\bf Acknowledgments}: LK thanks Robert K\"{o}nig for introducing Levin-Wen model to him, and Xiao-Gang Wen for a valuable comment on the physical meaning of defects of codimension 3. We thank John Preskill, Hector Bombin, Anton Kapustin and Yong-Shi Wu for many inspiring conversations.  This work was supported in part by NSF under Grant No.\ PHY-0803371. AK is also supported by ARO under Grant No. W911NF-09-1-0442. LK is supported by the Gordon and Betty Moore Foundation through Caltech's Center for the Physics of Information, and NSF under Grant No. PHY-0803371, the Basic Research Young Scholars Program and the Initiative Scientific Research Program of Tsinghua University, and NSFC under Grant No. 11071134.

\section{Toric code} \label{toric}

\begin{figure}[t] 
  \centerline{\includegraphics[scale=1.44]{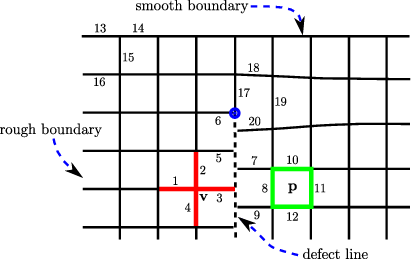}}
  \caption{Toric code with boundaries of two types and a defect line.}
  \label{toric-defect}
\end{figure}

Let us consider a variant of the ``toric code''~\cite{kitaev97} on a planar lattice with external boundaries and a defect line (domain wall), see Figure~\ref{toric-defect}.

The Hamiltonian is a sum of so-called \emph{stabilizer operators} taken with a minus sign. Such operators $X_k$ have eigenvalues $\pm 1$ and commute with each other. A (possibly non-unique) ground state satisfies the \emph{stabilizer conditions} $X_k|\psi\rangle=|\psi\rangle$ for all $k$. We use different notation for different types of stabilizer operators. In the bulk, the operators $A_{\mathbf{v}}$, $B_{\mathbf{p}}$ are defined for each vertex $\mathbf{v}$ and each plaquette $\mathbf{p}$. For example, for $\mathbf{v}$ and $\mathbf{p}$ in Figure~\ref{toric-defect}, $A_{\mathbf{v}}$ and $B_{\mathbf{p}}$ are defined as follows:
\begin{equation}
A_{\mathbf{v}} = \sigma_1^x\sigma_2^x\sigma_3^x\sigma_4^x, \quad\quad
B_{\mathbf{p}} = \sigma_8^z\sigma_{10}^z\sigma_{11}^z\sigma_{12}^z\,.
\end{equation}
(We also denote them by $A_{1,2,3,4}$ and $B_{8,10,11,12}$, respectively.) Boundary conditions will be discussed later.

We are interested in excited states that violate a small number of stabilizer conditions (compared to the system size). There are four superselection sectors of bulk excitations, $1, e, m, \epsilon$, where $e$ is an ``electric'' (actually, $\ZZ_2$) charge located on a vertex, and $m$ is a ``magnetic'' vortex on a plaquette.  $e$-particles and $m$-particles can be created in pairs at the ends of open strings.  Label $1$ denotes the trivial sector (no excitations at all or an even number of $e$ and $m$); $\epsilon$ can be obtained from $e$ and $m$ by fusion.  Thus, $e\times e=m\times m=1$ and $e\times m =\epsilon$. The four sectors form the unitary modular category $Z(\text{Rep}_{\mathbb{Z}_2})$, the center of the representation category 
$\text{Rep}_{\mathbb{Z}_2}$ of the group $\mathbb{Z}_2$.

There are two boundary types~\cite{bk98}, ``smooth'' and ``rough'', as depicted in Figure \ref{toric-defect}. The stabilizer operators on the boundaries are defined like this:
\begin{equation}
A_{13,14,15} = \sigma_{13}^x \sigma_{14}^x \sigma_{15}^x, \quad\quad 
B_{13,15,16} = \sigma_{13}^z \sigma_{15}^z \sigma_{16}^z\, .
\end{equation}
When a magnetic vortex approaches a smooth boundary, it disappears completely, but an electric charge can not pass and becomes a boundary excitation. Therefore, the excitations on the smooth boundary are $\{1,e\}$; they constitute the tensor category $\Rep_{\ZZ_2}$. Furthermore, the bulk excitations fuse into the boundary as 
\be  \label{eq:bulk-sbound}
1,m\mapsto 1,  \quad\quad e,\epsilon \mapsto e,
\ee
which gives the forgetful monoidal functor $F_{\text{smooth}}: Z(\Rep_{\ZZ_2})\to\Rep_{\ZZ_2}$. At the rough boundary, electric charges disappear, but magnetic vortices get stuck. As a result, the rough boundary excitations are $\{1,m\}$; they make the category $\Hilb_{\ZZ_2}$ of $\ZZ_2$-graded Hilbert spaces.\footnote{The grading has nothing to do with (anti)commutativity of the tensor product.  In fact, swapping the tensor factors in this context does not make sense because particles have a fixed order on the boundary.} The bulk-to-boundary map is 
\be \label{eq:bulk-rbound}
1,e\mapsto 1, \quad\quad m,\epsilon\mapsto m, 
\ee
which gives another monoidal functor, $F_{\text{rough}}: Z(\text{Rep}_{\mathbb{Z}_2})\rightarrow\Hilb_{\ZZ_2}$. Note that $\Hilb_{\ZZ_2}\cong\Rep_{\ZZ_2}$.

The dotted line (representing a 1d defect, or a domain wall) in Figure~\ref{toric-defect} does not carry any spins. The stabilizer operators on this line act on adjacent spins, e.g.,
\begin{equation}
C_{2,5,3|7}= \sigma_2^z \sigma_5^z \sigma_3^z \sigma_7^x, \quad 
D_{3|7,8,9}= \sigma_3^x \sigma_7^z \sigma_8^z \sigma_9^z \, . 
\end{equation}
Mathematically, such a defect line is represented by the category $\Hilb$ of Hilbert spaces, which is equipped with a structure of a $\Rep_{\ZZ_2}$-$\Rep_{\ZZ_2}$-bimodule (see Section \ref{sec:lw-defect}).

Let us see what happens when a bulk excitation approaches the defect. An electric charge located at $\mathbf{v}$ in Figure \ref{toric-defect} is represented by a state $|\xi\rangle$ such that $A_{\mathbf{v}}|\xi\rangle=-|\xi\rangle$, but all the other stabilizer conditions are satisfied. The application of $\sigma_3^z$ moves the charge to the defect line, breaking the condition involving $D_{3|7,8,9}$. If we further apply $\sigma_8^x$, that condition is restored and the quasiparticle becomes a magnetic vortex located at plaquette $\mathbf{p}$:
\[
B_{p}|\eta\rangle=-|\eta\rangle,\quad \text{where }
|\eta\rangle=\sigma_8^x\sigma_3^z|\xi\rangle\,.
\]
Similarly, a magnetic charge on the left-hand side of the defect line can be transformed to an excitation on the defect (breaking some $C$ condition) to an electric charge on its right-hand side. This process is clearly reversible.

Thus, the bulk quasiparticles can pass through the defect both ways according to this rule:
\begin{equation}  \label{eq:toric-tunnel}
1\leftrightarrow 1,\quad e\leftrightarrow m,\quad m\leftrightarrow e,\quad
\epsilon\leftrightarrow\epsilon.
\end{equation}
Mathematically, this situation is described by a pair of monoidal isomorphisms
\begin{equation} \label{eq:bulk-to-defect-toric}
Z(\Rep_{\ZZ_2}) \xrightarrow{L_{\Hilb}}
\calE_{\Hilb} \xleftarrow{R_{\Hilb}}
Z(\Rep_{\ZZ_2})
\end{equation}
where $\calE_{\Hilb}$ is the tensor category representing the excitations on the defect line.  Furthermore, the monoidal functor $T_{\Hilb}:=R_{\Hilb}^{-1}\circ L_{\Hilb}$ preserves the braiding, therefore it is an automorphism of the modular category $Z(\Rep_{\ZZ_2})$. This automorphism can be realized by moving quasiparticles along a loop encircling the defect line endpoint.

At the end of the defect line in Figure \ref{toric-defect}, there is another stabilizer operator 
\begin{equation}  \label{eq:Q}
Q=\sigma_{6}^x\sigma_{17}^y\sigma_{18}^z\sigma_{19}^z\sigma_{20}^z,
\end{equation}
which commutes with the stabilizer operators defined earlier. It represents the transport of an $\epsilon$-particle around the defect endpoint. Each choice of the eigenvalue ($+1$ or $-1$) of $Q$ determines a local superselection sector. We will show in Section \ref{sec:defect-cod-23} that such endpoints of defect lines can be viewed as generalized excitations, or defects of codimension 2. They are classified by distinct simple objects in a category of bimodule functors $\Fun_{\text{Rep}_{\mathbb{Z}_2}|\text{Rep}_{\mathbb{Z}_2}}(\text{Rep}_{\mathbb{Z}_2},\Hilb)$ (or $\Fun_{\text{Rep}_{\mathbb{Z}_2}|\text{Rep}_{\mathbb{Z}_2}}(\Hilb,\text{Rep}_{\mathbb{Z}_2})$ by a different convention). When a finite defect line contracts and disappears, the endpoints fuse into a bulk excitation. If both ends correspond to the same eigenvalue of $Q$, their fusion results in $1$ or $\epsilon$; if they are different, they fuse into $e$ or $m$ (this fusion rules were also found by Bombin~\cite{bombin}). In the mathematical language, such fusion corresponds to the composition of bimodule functors.

\section{Levin-Wen models} \label{levin-wen}

The toric code model is just a special case of a more general construction by Levin and Wen~\cite{lw-mod}. For an arbitrary unitary tensor category $\calC$, they defined a lattice model such that its bulk excitations are given by $Z(\calC)$, the monoidal center of $\calC$. We now outline this construction. 

By definition, a unitary category $\calC$ is equivalent to the category of formal sums, $X= \oplus_{i\in I} X^{(i)} \cdot i$, where $I$ is a finite set of inequivalent simple objects in $\calC$, and $X^{(i)}$ are arbitrary finite-dimensional Hilbert spaces. A morphism $f\in\Hom_{\calC}(X,Y)$ is just a collection of linear maps $f^{(i)}:X^{(i)}\to Y^{(i)}$. (Note that $\Hom_\calC(i, X)\cong X^{(i)}$.) The adjoint morphism $f^*:Y\to X$ is defined componentwise.

A \emph{unitary tensor category}  (UTC) is also equipped with an associative tensor product $\otimes$ and a unit object $\one\in\calC$, which is also called ``vacuum''. The associativity of the tensor product is expressed by isomorphisms $(X\otimes Y)\otimes Z \xrightarrow{\cong} X \otimes(Y\otimes Z)$ (the associator), and $\one \otimes X \cong X \cong X \otimes \one$. Moreover, each object $X$ in $\calC$ has a two-sided dual $\dual{X}$ satisfying certain properties that turn $\calC$ into a spherical category \cite{kitaev06}. For the most part we will assume that $\one$ is simple\footnote{A UTC with non-simple unit appears below in some supplementary construction. Levin-Wen models for such UTCs will be studied in \cite{kk}.}, i.e. $\one\in I$. 
  
For each UTC $\calC$, Levin and Wen defined a quantum model on a lattice (they called it a ``string-net model''). A ``lattice'' or ``string net'' is just an oriented planar graph with or without external edges, see Fig.~\ref{fig:lw-mod}. (More generally, one could consider graphs on oriented surfaces.) The planarity implies that the legs of each vertex are arranged into a clockwise cyclic order.  For simplicity, we assume that all vertices are trivalent.\footnote{Our definition of vertex spins works only for trivalent graphs. Some additional formalism is necessary to deal with general planar graphs~\cite{kk}.} Associated with the UTC and the the graph are two Hilbert spaces, $\calL\subseteq\calH$. The Levin-Wen Hamiltonian acts in the \emph{physical Hilbert space} $\calH$, which is the tensor product of spaces assigned to the graph edges and vertices. The subspace $\calL$ is spanned by \emph{diagrams} which describe compositions of elementary morphisms in $\calC$. It is a more natural construction from the mathematical point of view, whereas the definition of the Levin-Wen model allows some variations.

Basis vectors of both $\calL$ and $\calH$ are defined by labels that are assigned to each edge and each vertex of the graph. The edge labels are simple objects in $\calC$. A vertex label, or \emph{spin} is given by a choice of 3 indices $i_1,i_2,i_3\in I$ and a basis vector $\alpha$ in an associated morphism space. The basis vectors of $\calL$ correspond to \emph{stable labelings}, meaning that $i_1,i_2,i_3$ must match the labels on the incident edges, whereas the Levin-Wen space $\calH$ is spanned by all labelings. To determine which $\Hom$ space the vertex spin lives in, we first rotate the vertex to a standard position such that the incoming legs are on the bottom and the outgoing legs are on the top, then we read the morphism $\alpha$ from the bottom to the top. For example, if the three legs are oriented like this, 
\be  \label{eq:spin-label}
\raisebox{-20pt}{
  \begin{picture}(75,53)
   \put(10,8){\scalebox{1}{\includegraphics{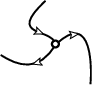}}}
   \put(0,8){
     \setlength{\unitlength}{.75pt}\put(0,0){
     \put(8,19)  {\scriptsize $ i $}
     \put(45,53)  {\scriptsize $ j $}
      \put(51,20)  {\scriptsize $ \alpha $}
     \put(70,-8)  {\scriptsize $ k $}
  }\setlength{\unitlength}{1pt}}
  \end{picture}}
  :=
 \raisebox{-20pt}{
  \begin{picture}(75,53)
   \put(10,8){\scalebox{1}{\includegraphics{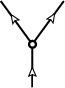}}}
   \put(0,8){
     \setlength{\unitlength}{.75pt}\put(0,0){
     \put(10,59)  {\scriptsize $ k $}
     \put(53,59)  {\scriptsize $ i $}
      \put(38,23)  {\scriptsize $ \alpha $}
     \put(33,-8)  {\scriptsize $ j $}
  }\setlength{\unitlength}{1pt}}
  \end{picture}}
\ee
then $\alpha$ is a basis vector in $V^{ki}_{j}=\Hom_\calC(j, k\otimes i)$. If all three legs are oriented outward, then $\alpha\in V^{ijk}=\Hom_\calC(\one, i\otimes j\otimes k)$. In the latter case, the cyclic order should be supplemented by a choice of the first leg to avoid ambiguity. All properties that we discuss are independent of the edge orientation. The direction of any single edge can be reversed if we also change its label $i$ to $\dual{i}$ and apply certain unitary operators to the spins at its endpoints.\footnote{The spin transformations involve an isomorphism between two variants of $\dual{i}$: the dual to the object associated with label $i$ and the object associated with the dual label. This isomorphism is defined up to a phase factor. However, the non-universal phases cancel each other if the two spins at the edge endpoints are stable.}

\begin{figure}
 \raisebox{-160pt}{
  \begin{picture}(160,160)
   \put(130,8){\scalebox{1.5}{\includegraphics{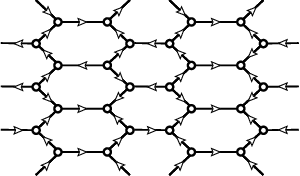}}}
   \put(130,8){
     \setlength{\unitlength}{.75pt}\put(-151,-235){
     \put(269,335)  {\scriptsize $ i $}
     \put(257,357)  {\scriptsize $ j $}
     \put(228,348)  {\scriptsize $ k $}
     \put(248, 331) {\scriptsize $\alpha$}
     }\setlength{\unitlength}{1pt}}
  \end{picture}}
\caption{\label{fig:lw-mod} An oriented planar graph with edge and vertex labels.}
\end{figure}
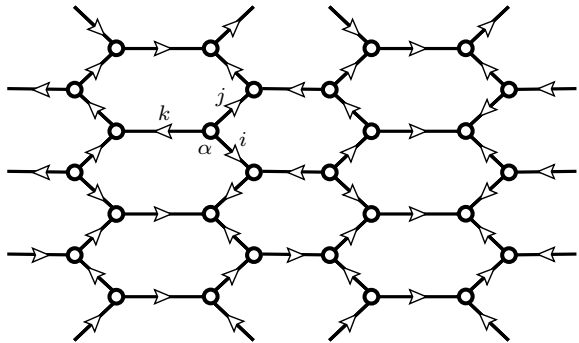

As mentioned above, planar graphs with stable labelings can be interpreted as diagrams that describe compositions of elementary morphisms in $\calC$. The axioms of a tensor category are thus represented by certain relations between diagrams. For example, the associator $(X\otimes Y)\otimes Z \xrightarrow{\cong} X\otimes(Y\otimes Z)$ and the unit isomorphisms $\one \otimes X \cong X \cong X\otimes \one$ should satisfy the pentagon relation and the triangle relation. Written in a basis, the associator is represented by so-called $F$-matrices, and the corresponding transformation of graphs is called an ``$F$-move'':
\begin{equation} \label{eq:fmatrix-C}
\figbox{1.0}{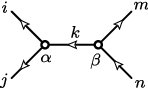} \quad=\,\,
\sum_{l}\sum_{\mu,\nu}\,
\langle l,\mu,\nu|F_{\,n}^{jim}|k,\alpha,\beta\rangle \,\,\,
\figbox{1.0}{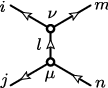}\quad,
\end{equation}
where $F_{\,n}^{jim}: \bigoplus_{k} V_{k}^{ji}\otimes V_{n}^{km} \to \bigoplus_{l}V_{n}^{jl}\otimes V_{l}^{im}$. Equality between diagrams means that they are evaluated to the same morphism (in this case, between $n$ and $j\otimes i\otimes m$). The pentagon equation for associators is equivalent to the pentagon identity for $F$-matrices. The triangle relation simply says that $F_{\,n}^{j\one m} = 1$ for a certain choice of basis.

In the previous discussion, we alluded to a Hermitian inner product on morphism spaces. It is defined as follows:
\begin{equation}\label{eq:inner_product}
\langle\eta|\xi\rangle\,=\,
\frac{1}{\sqrt{d_id_jd_k}}\,\,\figbox{1.0}{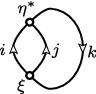}\qquad\quad
\text{for}\,\: \xi,\eta\in V^{ij}_{k},
\end{equation}
where $\eta^*\in V^{k}_{ij}$ is the adjoint morphism, and $d_X$ is the quantum dimension of object $X$. The normalization factor is chosen so as to satisfy two requirements: (i) it is symmetric and (ii) the associator $F_{\,n}^{jim}$ is unitary. Quantum dimensions also appear in the formula for the decomposition of the identity:
\begin{equation}\label{eq:decomp_id}
\figbox{1.0}{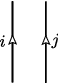}\,=\,\sum_{k\in I}\sum_{\alpha}
\sqrt{\frac{d_k}{d_id_j}}\,\figbox{1.0}{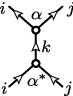}\,\,,
\end{equation}
where $\alpha$ runs over an orthonormal basis of $V^{ij}_{k}$.\smallskip

\begin{figure}[t]
\centerline{\begin{tabular}{@{}c@{\qquad}c@{\qquad}c@{\qquad}c@{}}
\includegraphics{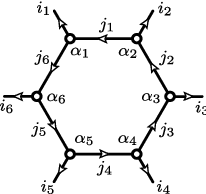}&
\includegraphics{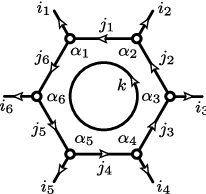}&
\includegraphics{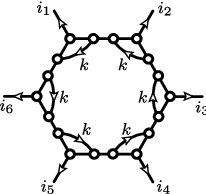}&
\includegraphics{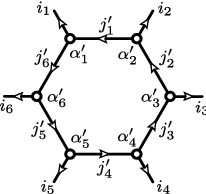}\\[5pt]
a) & b) & c) & d)
\end{tabular}}
\caption{The action of the plaquette operator $B_{\mathbf{p}}^k$:\newline
a) the initial state of the plaquette;\quad
b) a symbolic representation of the operator $B_{\mathbf{p}}^k$ applied to it;\newline
c) the loop is partially fused using Eq.~(\ref{eq:decomp_id}) (some labels and the overall factor are not shown);\newline
d) the corner triangles have been evaluated to trivalent vertices (summation over $j_p'$, $\alpha_q'$ is assumed).}\label{fig:Bpk}
\end{figure}

The Levin-Wen Hamiltonian is as follows: 
\begin{equation}  \label{ham}
H = \sum_{\mathbf{v}}(1-Q_{\mathbf{v}}) + \sum_\mathbf{p} (1-B_{\mathbf{p}}),
\qquad\quad \text{where}\quad
B_{\mathbf{p}}=\sum_{k\in I}\frac{d_k}{D^2}B_{\mathbf{p}}^k,\quad
D^2=\sum_{i\in I} d_i^2.
\end{equation}
The sums in the expressions for $H$ run over the vertices $\mathbf{v}$ and plaquettes $\mathbf{p}$ of the planar graph. The stabilizer operator $Q_{\mathbf{v}}$ acts on a label configuration (a basis vector in the physical space) $|\psi\rangle$ as follows: 
\[
Q_{\mathbf{v}} |\psi\rangle = \begin{cases}
    |\psi\rangle &   \text{if the vertex spin on $\mathbf{v}$ is stable}, \\
    0&   \text{otherwise}.
    \end{cases}
\]
The operator $B_{\mathbf{p}}^k$ acts as zero if some of the spins at the corners of plaquette $p$ are unstable. If all the spins at the corners are stable, then $B_{\mathbf{p}}^k$ creates a loop with label $k$, which is fused with the plaquette boundary using the rules of the tensor category, see Fig.~\ref{fig:Bpk} or the original paper~\cite{lw-mod}. The operators $Q_{\mathbf{v}}$ and $B_{\mathbf{p}}^k$ for different vertices and plaquettes commute with each other; $Q_{\mathbf{v}}$ and $B_{\mathbf{p}}$ are orthogonal projectors.

Ground states of the Levin-Wen model satisfy the stabilizer conditions
\begin{equation}
Q_{\mathbf{v}}|\psi\rangle=|\psi\rangle,\quad\,\,
B_{\mathbf{p}}|\psi\rangle=|\psi\rangle\qquad\,\,
\text{for all $\mathbf{v}$ and $\mathbf{p}$}.
\end{equation}
The first condition is simply the requirement that $|\psi\rangle\in\calL$. For a planar graph with fixed labels on its external edges, the ground space is isomorphic to the $\Hom$ space $V$ defined by those labels. The isometric embedding $V\to\calL$ is given by $D^{-p}\eval^{*}$, where $p$ is the number of plaquettes, and the map\, $\eval:\calL\to V$ corresponds to the evaluation of a diagram by the tensor category rules.

The excitations in the bulk were studied in \cite{lw-mod}. They are given by simple objects in $Z(\calC)$, the monoidal center of $\calC$. In this work, we will reinvestigate this problem in a more systematic way. In Section \ref{sec:lw-boundary}, we will identify boundary excitations with representations of a certain operator algebra and characterize them abstractly as module functors. In Section \ref{sec:lw-defect}, we will reduce a domain wall to an external boundary by folding the plane along the wall, then show that the bulk excitations are given by $Z(\calC)$ as a special case of wall excitations on the trivial domain wall.

\section{Gapped boundaries}  \label{sec:lw-boundary}

The Hamiltonian~(\ref{ham}) on a planar graph without external edges defines a Levin-Wen model with boundary. The labels on the boundary edges are still given by simple objects in $\calC$. We call such boundary type a $\calC$-boundary. 

Is that all the possible boundary types? The essential properties of the Levin-Wen model (e.g.\, the commutation between the operators $Q_{\mathbf{v}}$ and $B_{\mathbf{v}}^{s}$) follow from certain compatibility conditions of local data, in particular, the pentagon equation. Therefore, to find all boundary types, we need to define and solve the compatibility conditions in a neighborhood of the boundary.

\begin{figure} 
\centerline{\includegraphics{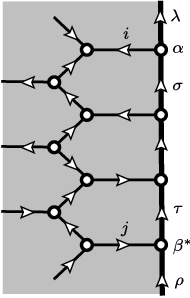}}
\caption{\label{fig:lw-boundary}
A neighborhood of the boundary in a Levin-Wen model.}
\end{figure}

A straightforward generalization (see Fig.~\ref{fig:lw-boundary}) involves a completely different set of labels $\lambda,\sigma,\tau,\rho$ on boundary edges. Those labels may correspond to simple objects in another unitary category $\calM$, which is not necessarily monoidal. Since there is no duality in general for objects in $\calM$, we can not change the orientation of boundary edges arbitrarily. Let us assume that all the boundary is oriented in a single direction so that the bulk stays on its left as shown in Figure~\ref{fig:lw-boundary}. The vertices on the boundary are labeled by basis vectors in some Hilbert spaces, e.g.\ $\alpha\in V^{i\lambda}_{\sigma}$,\, $\beta^*\in V^{\tau}_{j\rho}$ (these two cases differ in the orientation of the non-boundary edge). If $i=\one$, then only $\lambda=\sigma$ is allowed, and the corresponding Hilbert space is one-dimensional. Again, we should allow an $F$-move:
\begin{equation} \label{fmatrix-M}
\figbox{1.0}{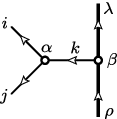}\quad=\,\,\,
\sum_{\sigma}\sum_{\mu,\nu}\,
\langle \sigma,\mu,\nu|F_{\,\rho}^{ji\lambda}|k,\alpha,\beta\rangle\,\,\,
\figbox{1.0}{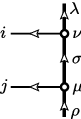}\quad,
\end{equation}
where $F_{\,\rho}^{ji\lambda}: \bigoplus_{k} V_{k}^{ji}\otimes V_{\rho}^{k\lambda}\to \bigoplus_{\sigma}V_{\rho}^{j\sigma}\otimes V_{\sigma}^{i\lambda}$. The matrices $F_{\,\rho}^{ji\lambda}$ should also satisfy the pentagon and the triangle identities, the later of which amounts to $F_{\,\rho}^{j\one\lambda}$ being equal to $1$ in a certain basis. 

Mathematically, a category $\M$ equipped with the above structures is called a \emph{left $\calC$-module}. Namely, there is an action $\otimes: \calC \times \calM \rightarrow \calM$ of $\calC$ on $\calM$ satisfying some axioms: the associativity of the action, the unit property and certain coherence properties, which correspond exactly to the pentagon and the triangle equations. In this new language, the Hilbert space of the vertex spin is  $V_{\sigma}^{i\lambda}=\Hom_\calM(\sigma, i\otimes\lambda)$ or $V_{i\lambda}^{\sigma}=\Hom_\calM(i\otimes\lambda, \sigma)$ (depending on the edge orientation), and we have $V_{\sigma}^{\one\lambda} \cong V_{\one\lambda}^{\sigma} \cong \delta_{\sigma\lambda} \mathbb{C}$. We say that these algebraic data define a ${}_\calC\calM$-boundary. To ensure that the quantum model is stable to perturbations, we will assume that the module $\calM$ is \emph{indecomposable}, i.e.\ for every pair of simple objects $\sigma,\lambda\in\calM$ there is some $i\in\calC$ such that $V_{\sigma}^{i\lambda}\not=0$.

The string-net Hamiltonian for the system with an $\calM$-boundary is defined again by formula~(\ref{ham}). Like in the usual Levin-Wen model, the operator $B_{\mathbf{p}}^k$ inserts a loop that is labeled by a simple object $k$ of the tensor category $\calC$. The only new thing is that the action of $B_{\mathbf{p}}^k$ on a plaquette $\mathbf{p}$ adjacent to the boundary involves both the $F$-matrices in $\calC$ and the $F$-matrices defined in equation~(\ref{fmatrix-M}). The operators $Q_{\mathbf{v}}$ and $B_{\mathbf{p}}^k$ are all mutually commutative, and $B_{\mathbf{p}}:=\sum_k \frac{d_k}{D^2} B_{\mathbf{p}}^k$ are projectors.

Before discussing concrete properties of this model, let us outline some pertinent mathematical structure. As already mentioned, a left $\calC$-module does not have intrinsic duality. However, one can define a dual of $X\in\calM$ as an object in the opposite category, $\calM^\op$. It consists of the same, or one-to-one related objects as $\calM$, but all arrows are reversed. Thus, each $X\in\calM$ is associated with its dual, $\dual{X}\in\calM^\op$, and $\Hom_{\dual{\calM}}(\dual{Y},\dual{X})$ is canonically identified with $\Hom_{\calM}(X,Y)$. We can now allow oppositely oriented boundary edges with labels in $\calM^\op$. A vertex with an incoming edge $i$ from the bulk and two outgoing edges $\lambda$, $\dual{\sigma}$ on the boundary carries a spin in $V_{i}^{\lambda\dual{\sigma}}:=V_{i\sigma}^{\lambda}$. Such spins may be interpreted as $\calC$-morphisms between $i$ and $\lambda\otimes\dual{\sigma}$, where the latter is defined by
\begin{equation}\label{eq:XYbar}
X\otimes\dual{Y}
:=\bigoplus_{i\in I}\Hom_{\calM}(i\otimes Y,\,X)\cdot i
\qquad \text{for}\,\: X,Y\in\calM.\qquad\qquad
\vcenter{\hbox{\includegraphics{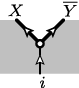}}}
\end{equation}
The picture illustrates a restriction on the $\Hom$ space: the thin line can only join the thick one from the shaded side. An analogous definition for $\dual{Y}\otimes X$ does not work. Nevertheless, $\dual{Y}\otimes X$ can be defined as an object in a certain UTC $\calD=\calC_{\calM}^\vee$ (see below). The four categories, $\calC$, $\calD$, $\calM$, $\calM^\op$ form a so-called Morita context~\cite{mueger}. The objects of these categories and their formal sums constitute a UTC $\calB$ with non-simple unit, $\one_{\calB}=\one_{\calC}\oplus\one_{\calD}$. This construction implies a canonical duality in $\calB$, which is pivotal and spherical (in a generalized sense). Thus, the quantum dimensions are defined for objects in $\calB$ in general and for boundary labels in particular.\smallskip

\begin{figure}
\centerline{\begin{tabular}{@{}c@{\qquad\quad}c@{\qquad}c@{\qquad}c@{}}
\includegraphics{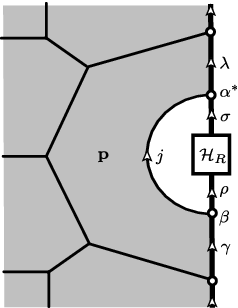}&
\includegraphics{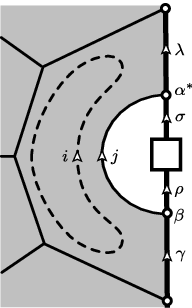}&
\includegraphics{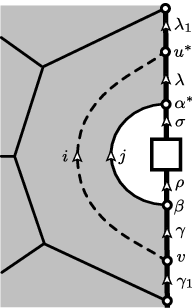}&
\includegraphics{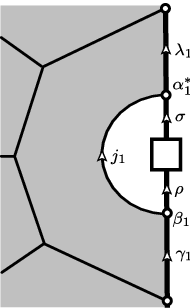}\\[5pt]
a) & b) & c) & d)
\end{tabular}}
\caption{Boundary excitations (the unexcited part of the lattice is shown in gray):\newline
a) the labels characterizing an excited state;\quad
b) operator $B_{\mathbf{p}}$ acts on the adjacent plaquette;\newline
c) the loop is partially fused;\quad d) the loop is completely fused.}\label{fig:edge-alg-action}
\end{figure}

Now we will study boundary excitations. Informally, an excitation is a small region $R$ near the lattice boundary where some of the stabilizer conditions are removed or altered. It may also contain some extra degrees of freedom that are not present in the original model. For simplicity, let us assume that the region $R$ is separated from the rest of the graph by a single edge,\footnote{If that is not the case, we can use $F$-moves to shrink a separating path to a single edge. Thus we obtain an equivalent model on a different lattice.} see Figure~\ref{fig:edge-alg-action}. Thus, excited states belong to the tensor product $\calH_{\mathrm{ext}}\otimes A\otimes\calH_{R}$. The three factors are as follows: $\calH_{R}$ represents the interior of $R$; the space
\begin{equation}\label{eq:algA}
A = \bigoplus_{j,\lambda,\sigma,\gamma,\rho}
V^{\lambda}_{j\sigma} \otimes V^{j\rho}_{\gamma}\qquad\qquad
\left(\text{basis vectors}\quad
\alpha^*\otimes\beta=\,\figbox{1.0}{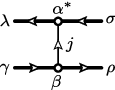}\,\right)
\end{equation}
includes the separating edge $j$, its endpoints, and the adjacent edges; $\calH_{\mathrm{ext}}$ describes the exterior, i.e.\ the unexcited part. All the stabilizer operators $Q_{\mathbf{v}}$, $B_{\mathbf{p}}$ acting in $\calH_{\mathrm{ext}}\otimes A$ are in place. In addition, the excited states may be constrained by some projector $P_R$ that acts in $A\otimes\calH_{R}$ and commutes with $B_{\mathbf{p}}$ on the adjacent plaquette. Hence, the set of excited states is characterized by the subspace $E=\mathrm{Im}\,P_R\subseteq A\otimes\calH_{R}$. We set aside some ambiguities in this definition. For example, the region $R$ can be extended to a larger region $R'$ so that the pair $(\calH_{R},E)$ and the new pair $(\calH_{R'},E')$ describe the same excitation. This and other subtleties will be addressed in~\cite{kk}.\enlargethispage{0.3cm}

The key observation is that the subspace $\calH_{\mathrm{ext}}\otimes E$ is invariant under the action of $B_{\mathbf{p}}$ on the whole system. This action is depicted in Figure~\ref{fig:edge-alg-action}. Let us decompose $B_{\mathbf{p}}$ as $\sum_{t}Q^{(\mathrm{ext})}_{t}\otimes Q^{(A)}_{t}$, where $Q^{(\mathrm{ext})}_{t}$ and $Q^{(A)}_{t}$ act in the respective spaces and are indexed by $t=(i,\lambda_1,\lambda,\gamma_1,\gamma,u,v)$. The operators $Q^{(A)}_{t}$ correspond to fusing the dashed line with the separating edge (the transition between Fig.~\ref{fig:edge-alg-action}c and~\ref{fig:edge-alg-action}d). One can show that the corresponding operators $Q^{(\mathrm{ext})}_{t}$ are linearly independent. Thus, the subspace $E\subseteq A\otimes\calH_{R}$ is invariant under the algebra generated by $Q^{(A)}_{t}$. As a linear space, this algebra is isomorphic to $A$ defined by~(\ref{eq:algA}). The multiplication rule is as follows:
\begin{equation}
\figbox{1.0}{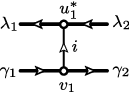}\quad\mult\quad\figbox{1.0}{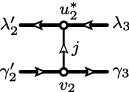}
\quad=\quad \delta_{\lambda_2\lambda'_2}\delta_{\gamma_2\gamma'_2}\quad
\figbox{1.0}{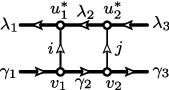}
\end{equation}
where the last graph is reduced to a linear span of graphs in $A$ by applying $F$-moves~(\ref{fmatrix-M}) twice and removing bubbles (c.f. \cite{lw-mod}, Eq.~(6)). The unit of $A$ is given by $\oplus_{\lambda, \gamma} \id_\lambda\otimes \id_\gamma$. Since this algebra acts on $E$, we can identify a boundary excitation with an $A$-module (i.e.\ a representation of $A$).

\begin{figure}  
\centerline{$\displaystyle
\figbox{1.0}{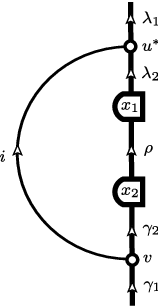}
\quad\,\,=\quad\sum_{\tau,w}\,\sqrt{\frac{d_\tau}{d_id_\rho}}\,\,
\figbox{1.0}{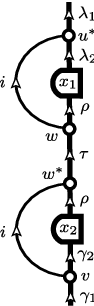}
$}
\caption{\label{fig:fusion-comulti}
Simultaneous action of algebra $A$ on two excitations. The sum is over all simple objects $\tau\in\calM$ and basis vectors $w\in V^{i\rho}_{\sigma}$. On the one hand, this equation is a diagrammatic identity that follows from a variant of~(\ref{eq:decomp_id}). On the other hand, it defines a comultiplication on $A$.}
\end{figure}

Structures on $A$ can be enriched to give a bialgebra with a comultiplication $\Delta: A \to A \otimes A$. The comultiplication defines the fusion of two boundary excitations $E_1$, $E_2$ (or equivalently, the tensor product $E_1\otimes E_2$ as an $A$-module). More precisely, in terms of Sweedler's notation,\footnote{The expression $a^{(1)}\otimes a^{(2)}$ involves an implicit sum and should be understood as $\sum_{t}a^{(1)}_t\otimes a^{(2)}_t$.} $\Delta(a) = a^{(1)}\otimes a^{(2)}$, the simultaneous action of $a\in A$ on two excitations is defined as follows:
\be  \label{eq:comulti-fusion}
a \cdot (x_1 \otimes x_2) :=  (a^{(1)} \cdot x_1) \otimes (a^{(2)} \cdot x_2).
\ee
In our case, equation (\ref{eq:comulti-fusion}) is explicitly depicted in Figure \ref{fig:fusion-comulti}.

Actually, $A$ is a $C^\ast$ weak Hopf algebra rather than just a bialgebra. Weak Hopf algebras were introduced by Bohm and Szlachanyi~\cite{bsz}; the axioms in a concise form can be found in~\cite{nikshych}. The connection between weak Hopf algebras and module categories was established by Ostrik~\cite{ostrik}. In our setting, there is an additional $C^\ast$ structure related to unitarity. The star, the antipode, and all other operations are given below. To make the definitions more symmetric, we have introduced additional factors in the formulas for the multiplication and comultiplication. This is just a basis change: each basis vector in equation~(\ref{eq:algA}) is multiplied by $(d_{\sigma}d_{\rho})^{-1/2}$. (For notational convenience, we sometimes assign labels to parts of diagrams, which are enclosed in dotted lines.)

\begin{align}
&\eqtitle{Multiplication}\quad&&
\figbox{1.0}{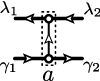}\quad\mult\quad\figbox{1.0}{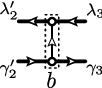}
\quad=\quad
\frac{\delta_{\lambda_2\lambda_2'}\,\delta_{\gamma_2\gamma_2'}}
{\sqrt{d_{\lambda_2}d_{\gamma_2}}}\quad
\figbox{1.0}{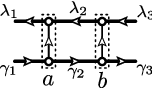}
\\[8pt]
&\eqtitle{Unit}\quad&&
e\quad=\quad
\sum_{\sigma,\rho}\,\sqrt{d_{\sigma}d_{\rho}}\:\figbox{1.0}{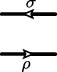}
\\[3pt]\nonumber\\
&\eqtitle{Comultiplication}\quad&&
\Delta\hskip1em\underbrace{\hskip-1em\left(\figbox{1.0}{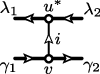}
\right)\hskip-1em}_{\textstyle a}\hskip1em
\quad=\quad
\sum_{\tau,\rho,w}\,
\underbrace{\sqrt{\frac{d_{\tau}d_{\rho}}{d_i}}\:
\figbox{1.0}{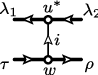}}_{\textstyle a^{(1)}}\,\:\otimes\,\:
\underbrace{\figbox{1.0}{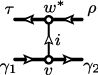}}_{\textstyle a^{(2)}}
\\[8pt]
&\eqtitle{Counit}\quad&&
\varepsilon\left(\figbox{1.0}{mult1}\right)
\quad=\quad
\frac{\delta_{\lambda_1\gamma_1}\delta_{\lambda_2\gamma_2}}
{d_{\lambda_1}d_{\lambda_2}}\quad
\figbox{1.0}{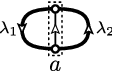}
\\[8pt]
&\eqtitle{Antipode}\quad&&
S\left(\figbox{1.0}{comult1}\right)
\quad=\quad
\sqrt{\frac{d_{\gamma_1}d_{\lambda_2}}{d_{\lambda_1}d_{\gamma_2}}}\quad
\figbox{1.0}{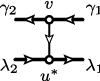}\qquad (\text{$180^{\circ}$ rotation})
\\[8pt]
&\eqtitle{Star}\quad&&
\left(\figbox{1.0}{comult1}\right)^{\!*}
\quad=\quad
\figbox{1.0}{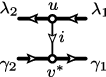}\hskip2.5cm (\text{horizontal flip})
\end{align}
\medskip

In the new basis, the action of the algebra $A$ on a representation space $E=\bigoplus_{\sigma,\rho}E^{\sigma}_{\rho}$ can be written symbolically:
\begin{equation}\label{eq:Aact}
\figbox{1.0}{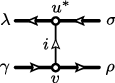} \,\,\cdot\,\, \figbox{1.0}{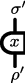} \quad=\quad \frac{\delta_{\sigma\sigma'}\,\delta_{\rho\rho'}}
{\sqrt{d_{\sigma}d_{\rho}}}\,\,
\figbox{1.0}{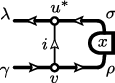}
\end{equation}
However, the box containing $x$ is not part of the ${}_\calC\calM$ calculus, so the exact meaning of the diagram on the right is not clear yet. We will see that the box can be interpreted as a new type of vertex in some extended diagrammatic calculus.

Boundary excitations can also be described using the mathematical notion of a \emph{unitary $\calC$-module functor}. It is a functor $\Phi:\calM\to\calM$ together with a natural isomorphism $F^{(\Phi)}$ that defines the commutation of $\Phi$ with the action of $\calC$ on $\calM$. Specifically, for each $X\in\calC$ and $Y\in\calM$ there are unitary isomorphisms $F^{(X,Y,\Phi)}:\Phi(X\otimes Y)\to X\otimes\Phi(Y)$. In addition to naturality, they satisfy certain other conditions that are similar to the pentagon and triangle equations \cite{ostrik}. To emphasize this analogy, let us replace the standard notation for the application of functor, $\Phi(X)$ with $X\otimes\Phi$. Then $F^{(X,Y,\Phi)}:(X\otimes Y)\otimes\Phi\to X\otimes (Y\otimes\Phi)$ is just a form of associator.

Written in a basis, the functor $\Phi$ is given by the linear spaces $V^{\sigma\Phi}_{\rho}=\Hom_{\calM}(\rho,\,\sigma\otimes\Phi)$. The Hermitian inner product between $x,y\in V^{\sigma\Phi}_{\rho}$ is defined by the equation $y^*x=\langle y|x\rangle\id_{\rho}$ (up to a normalization factor that is not important at the moment). The associator and its inverse can be represented in a matrix form, where $x$ and $y$ run over orthonormal bases of $V^{\sigma\Phi}_{\rho}$ and $V^{\lambda\Phi}_{\gamma}$, respectively:
\begin{align}
\figbox{1.0}{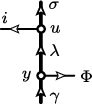}\,\,\,&=\,\,\,
\sum_{\rho}\sum_{x,v}\,
\langle\rho,x,v|F^{i\sigma\Phi}_{\gamma}|\lambda,y,u\rangle\,\,
\figbox{1.0}{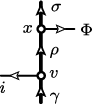}\,\,,
\label{eq:Fmatr12}\\[8pt]
\figbox{1.0}{Fmatr2}\,\,\,&=\,\,\,
\sum_{\lambda}\sum_{y,u}\,
\langle\lambda,y,u|(F^{i\sigma\Phi}_{\gamma})^{-1}|\rho,x,v\rangle\,\,
\figbox{1.0}{Fmatr1}\,\,.
\label{eq:Fmatr21}
\end{align}
Now, let us attach $u^*$ on the top of both graphs in~(\ref{eq:Fmatr21}):
\[
\figbox{1.0}{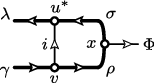}\quad
=\quad\sqrt{\frac{d_id_\sigma}{d_\lambda}}\,\sum_{y}\,
\langle\lambda,y,u|(F^{i\sigma\Phi}_{\gamma})^{-1}|\rho,x,v\rangle\,\,
\figbox{1.0}{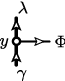}\,\,.
\]
The left-hand side looks like the last diagram in equation~(\ref{eq:Aact}), which is understood as an action of algebra $A$ on the Hilbert space $E=\sum_{\sigma,\rho}V^{\sigma\Phi}_{\rho}$. We have just expressed this action in terms of $F$-matrices. Another expression can be obtained by attaching $v$ on the bottom of~(\ref{eq:Fmatr12}). These two methods yield the following results:
\begin{align}
\figbox{1.0}{mult1-y}\,\,\cdot\,\, \figbox{1.0}{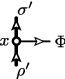}
\quad&=\quad
\sqrt{\frac{d_i}{d_{\lambda}d_{\rho}}}\,
\delta_{\sigma\sigma'}\,\delta_{\rho\rho'}\,\sum_{y}\,
\langle \lambda,y,u|(F^{i\sigma\Phi}_{\gamma})^{-1}|\rho,x,v\rangle\,\,
\figbox{1.0}{mult6-y}\,\,,\\[8pt]
\left(\figbox{1.0}{mult1-y}\right)^{\!*} \,\,\cdot\,\, \figbox{1.0}{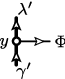} \quad&=\quad
\sqrt{\frac{d_i}{d_{\lambda}d_{\rho}}}\,
\delta_{\lambda\lambda'}\,\delta_{\gamma\gamma'}\,\sum_{x}\,
\langle \rho,x,v|F^{i\sigma\Phi}_{\gamma}|\lambda,y,u\rangle\,\,
\figbox{1.0}{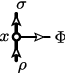}\,\,.
\end{align}
Note that the unitarity of $F^{i\sigma\Phi}_{\gamma}$ is equivalent to the condition $\bigl\langle y\big|(a\cdot x)\bigr\rangle =\bigl\langle(a^*\cdot y)\big|x\bigr\rangle$. Using the pentagon and the triangle equations, one can show that $(a\bullet b)\cdot x=a\cdot(b\cdot x)$ and $e\cdot x=x$. Thus, any unitary $\calC$-module functor from $\calM$ to $\calM$ defines a unitary representation of algebra $A$. The converse is also true.

Let us denote the category of unitary $\calC$-module functors from $\calM$ to $\calM$ by $\Fun_\calC(\calM, \calM)$, or $\calC_\calM^\vee$ for brevity. It is a unitary category whose simple objects correspond to irreducible representation of the algebra $A$, and thus describe superselection sectors of boundary excitations. Furthermore, $\calC_\calM^\vee$ comes with a tensor product given by the composition of functors in the ``opposite'' order, i.e.\ $(\Phi\otimes\Psi)(X):=\Psi(\Phi(X))$. This rule is consistent with our previous notation, $X\otimes\Phi:=\Phi(X)$ and simply says that $X\otimes(\Phi\otimes\Psi)=(X\otimes\Phi)\otimes\Psi$. Written in the basis of simple objects, this identity gives a new set of $F$-matrices. The duality in $\calC_\calM^\vee$ is given by the notion of adjoint functor.\footnote{For functors between unitary categories, the left and right adjoint functors are equivalent.} By definition, $V^{\sigma\dual{\Phi}}_{\rho} =\Hom_{\calM}(\rho,\,\sigma\otimes\dual{\Phi}) =\Hom_{\calM}(\rho\otimes\Phi,\,\sigma) =V^{\sigma}_{\rho\Phi}$, and the last space is equal to $\bigl(V_{\sigma}^{\rho\Phi}\bigr)^{*}$.

Thus, $\calC$ and $\calC_\calM^\vee$ are unitary tensor categories, and $\calM$ is a $\calC$-$\calC_\calM^\vee$ bimodule. Similarly, $\calM^\op$ is $\calC_\calM^\vee$-$\calC$ bimodule. If $X,Y\in\calM$, then $X\otimes\dual{Y}\in\calC$ is given by equation~(\ref{eq:XYbar}). We can also define $\dual{Y}\otimes X\in\calC_\calM^\vee$ as the $\calC$-module functor that takes each $Z\in\calM$ to $(Z\otimes\dual{Y})\otimes X$. The whole structure is a bicategory with two $0$-cells, the objects of $\calC$, $\calC_\calM^\vee$, $\calM$, $\calM^\op$ being $1$-cells. This particular type of bicategory is called \emph{Morita context} and defines a \emph{Morita equivalence} between the tensor categories $\calC$ and $\calC_\calM^\vee$ \cite{mueger}. In the diagrammatic calculus, we allow all planar graphs with plaquettes of two colors. The edges separating gray from white are labeled by objects in $\calM$ or $\calM^\op$ (depending on the orientation); the edges with gray or white on both sides are labeled by objects in $\calC$ or $\calC_\calM^\vee$, respectively.

\begin{figure}
\centerline{\begin{tabular}{@{}c@{\qquad\quad}c@{\qquad\quad}c@{}}
\includegraphics{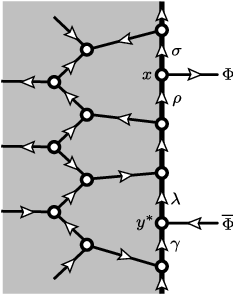}&
\includegraphics{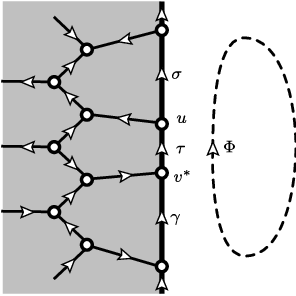}&
\includegraphics{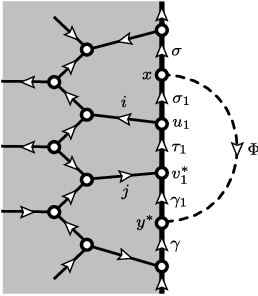}\\[5pt]
a) & b) & c)
\end{tabular}}
\caption{Interpretation of some extended diagrams in terms of the Levin-Wen model:\newline
a) a modified model with two special vertices that represent excitations;\newline
b) string operator $W_\Phi$ acting on the original model;\newline
c) the resulting state (the summation over $\sigma_1,\tau_1,\gamma_1$ is assumed).}\label{fig:stringop}
\end{figure}

What is the meaning of such extended diagrams in terms of the Levin-Wen model? Let us give a partial interpretation. Boundary vertices with external edges labeled by $\Phi\in\calC_\calM^\vee$ (as in Figure~\ref{fig:stringop}a) represent excitations.\footnote{From the physical point of view, excitations correspond to different subspaces $\calL'$ in a fixed physical space $\calH$. In this description, we have actually changed $\calH$ for the mathematical convenience.} Such a vertex carries the Hilbert space $E=\sum_{\sigma,\rho}V^{\sigma\Phi}_{\rho}$ on which the algebra $A$ acts; this action is used in the definition of the operators $B_{\mathbf{p}}^k$ on the adjacent plaquette. We can also construct an operator $W_{\Phi}$ (similar to the string operators of Levin and Wen~\cite{lw-mod}) which creates a pair of excitations from the ground state of the original model. To this end, consider a loop of $\Phi$-labeled string and fuse part of it with the boundary using the extended diagrammatic calculus, see Figure~\ref{fig:stringop}b,c. This process is equivalent to the application of functor $\Phi$ to the edge labels (i.e.\ objects of $\calM$) and vertex spins (morphisms); the resulting objects are then decomposed into simple ones, producing a linear combination of diagrams. The remainder of the string is not part of the model and can be simply erased. A more natural interpretation of the extended diagrammatic calculus is a model with ``transparent domain walls'', see the next section.

\section{Domain walls between phases} \label{sec:lw-defect}

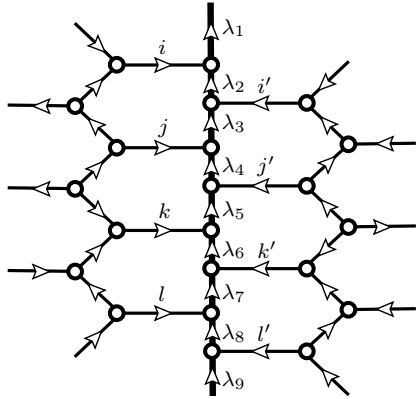
\begin{figure}
 \raisebox{-160pt}{
  \begin{picture}(160,160)
   \put(140,8){\scalebox{1.5}{\includegraphics{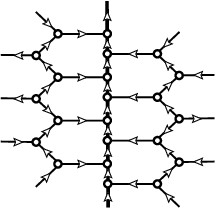}}}
   \put(140,8){
     \setlength{\unitlength}{.75pt}\put(-151,-235){
     \put(228,410)  {\scriptsize $ i $}
     \put(228,370)  {\scriptsize $ j $}
     \put(228,327)  {\scriptsize $ k $}
     \put(228,285)  {\scriptsize $ l $}
     \put(260,420)  {\scriptsize  $\lambda_1 $}
      \put(260,392)  {\scriptsize $\lambda_2 $}
     \put(260,372)  {\scriptsize $ \lambda_3 $}
      \put(260,350)  {\scriptsize $ \lambda_4 $}
     \put(260,328)  {\scriptsize $ \lambda_5 $}
     \put(260,308)  {\scriptsize $ \lambda_6$}
     \put(260,286)  {\scriptsize $ \lambda_7$}
      \put(260,266)  {\scriptsize $ \lambda_8 $}
       \put(260,242)  {\scriptsize $ \lambda_9 $}
     \put(278,390)  {\scriptsize $ i' $}
     \put(278,350)  {\scriptsize $ j' $}
     \put(278,307)  {\scriptsize $ k' $}
     \put(278,265)  {\scriptsize $ l' $}
     }\setlength{\unitlength}{1pt}}
  \end{picture}}
\caption{\label{lw-mod-defect}
A neighborhood of a defect line between two topological phases, where $i, j,k,l\in \calC$, $\lambda_1, \dots, \lambda_9 \in \M, i',j',k',l'\in \calD$.}
\end{figure}

Consider a planar lattice separated by a domain wall (defect line), which is depicted as the thicker line in Figure \ref{lw-mod-defect}. The labels $i,j,k,l$ and $i', j', k', l'$ are simple objects in unitary tensor categories $\calC$ and $\calD$ respectively,  and $\lambda_1, \dots \lambda_9$ come from another label set and can be viewed as simple objects in another unitary category $\calM$. The spins labels for the vertices on the wall are given by the basis vectors of $\Hom$-spaces in $\calM$. 

By folding the system along the domain wall, we can reduce the domain wall between two lattices associated with $\calC$ and $\calD$ to an external boundary of a doubled bulk system $\calC \boxtimes \calD^\opt$, where $\boxtimes$ is Deligne's tensor product \cite{de} and $\calD^\opt$ is the same category as $\calD$ but with the opposite tensor product $\mathop{\otimes}\limits^{\leftrightarrow}$, i.e. $X\mathbin{\mathop{\otimes}\limits^{\leftrightarrow}} Y = Y\otimes X$. By previous results, for a consistent model of string-net condensation, $\calM$ should be a left $\calC \boxtimes \calD^\opt$-module, or equivalently, a $\calC$-$\calD$-bimodule. We will refer to such domain wall a ${}_\calC\calM_\calD$-wall or a $\calM$-defect line. For example, the domain wall in the toric code (as shown in Figure~\ref{toric-defect}) is a $\Hilb$-defect line, where $\Hilb$ is considered as a $\text{Rep}_{\mathbb{Z}_2}$-$\text{Rep}_{\mathbb{Z}_2}$-bimodule. Conversely, the Levin-Wen model with a single ${}_\calC\calM$-boundary is just a special case of a domain wall between the $\calC$-bulk phase and the empty phase. The empty phase is described by the UTC $\Hilb$ with a single simple object, $\one$. Any left $\calC$-module $\calM$ is automatically a $\calC$-$\Hilb$-bimodule. In other words, an ${}_\calC\calM$-boundary is just a ${}_\calC\calM_{\Hilb}$-wall. 

By the results of Section \ref{sec:lw-boundary}, excitations on a domain wall are given by $\calC\boxtimes\calD^\opt$-module functors or, equivalently, $\calC$-$\calD$-bimodule functors from $\calM$ to $\calM$. We denote the category of all such functors by $\Fun_{\calC|\calD}(\calM, \calM)$, which is again a unitary tensor category \cite{eno05}. In the special case $\calD=\calC$, a trivial example of a domain wall is $\calM=\calC$ as a $\calC$-$\calC$-bimodule. The excitations on this trivial wall, $\Fun_{\calC|\calC}(\calC, \calC)$ are nothing but the bulk excitations. Mathematically, it is known that the monoidal center $Z(\calC)$ (defined independently) is equivalent to $\Fun_{\calC|\calC}(\calC, \calC)$ as braided tensor categories \cite{eo}. In this work, we simply define $Z(\calC)$ to be $\Fun_{\calC|\calC}(\calC, \calC)$. 

If we put an ${}_\calC \calM_\calD$-wall and a ${}_\calD \calN_\calE$-wall $\calN$ next to each other, viewed from far away, they fuse into a single wall given by $\calM \boxtimes_\calD \calN$, which is a $\calC$-$\calE$-bimodule defined by Tambara \cite{tambara}. A couple of cases of such domain wall fusion are especially interesting. First, $\calC \boxtimes_\calC \calM \cong \calM \cong \calM \boxtimes_\calD \calD$. Another case is when $\calC$ and $\calD$ are Morita equivalent. We have previously described this notion by constructing $\calD$ from $\calC$. The abstract definition is this: $\calC$ and $\calD$ are Morita equivalent if there exists a $\calC$-$\calD$-bimodule $\calM$ and a $\calD$-$\calC$-bimodule $\overline{\calM}$ such that $\calM\boxtimes_\calD \overline{\calM} \cong \calC$ and $\overline{\calM} \boxtimes_\calC \calM\cong \calD$ \cite{enom09}. Actually, $\overline{\calM}$ can be chosen to be $\calM^\op$ (the opposite category of $\calM$), and $\calD \cong \calM^\op \boxtimes_\calC \calM \cong \calC_\calM^\vee$. Thus, the construction from Section~\ref{sec:lw-boundary} covers the general case of Morita equivalence, and any UTC $\calD$ that is Morita equivalent to $\calC$ via $\calM$ describes the boundary excitations on an $\calM$-boundary of the ${}_{\calC}\calM$ Levin-Wen model. What we now consider is a continuation of that model past the boundary, which becomes a domain wall between $\calC$ and $\calD$ regions. Such a wall is called \emph{invertible}, or \emph{transparent}, because the bimodule $\calM$ is invertible with respect to the tensor product, and the bulk quasiparticles can ``tunnel'' through the wall as discussed below. The physical meaning of $\calM^\op$ is rather clear: it is an anti-domain wall of $\calM$. When we move domain walls $\calM$ and $\calM^\op$ close to each other, they annihilate.

When bulk excitations approach a domain wall, they fuse with it and become wall excitations. So we have two maps from bulk excitations on the two sides to wall excitations. We are ready to make these two maps mathematically explicit. Namely, they are the monoidal functors $$
Z(\calC) \xrightarrow{L_\calM} \Fun_{\calC|\calD}(\calM, \calM)  \xleftarrow{R_\calM} Z(\calD)
$$ 
given by
\bea \label{eq:L-bulk-to-defect}
L_\calM: \,\,\, (\calC \xrightarrow{\calF} \calC) &\longmapsto& (\calM \cong \calC \boxtimes_\calC \calM \xrightarrow{\calF \boxtimes_\calC \id_\calM} \calC \boxtimes_\calC \calM \cong \calM) \\
R_\calM: \,\,\, (\calD \xrightarrow {\calG} \calD) &\longmapsto& (\calM \cong \calM \boxtimes_\calD \calD \xrightarrow{\id_\calM \boxtimes_\calD \calG} \calM \boxtimes_\calD \calD \cong \calM) \nonumber
\eea
where $\calF$ is a $\calC$-$\calC$-bimodule functor, and $\calG$ is a $\calD$-$\calD$-bimodule functor. Notice that these maps exactly coincide with our physical intuition. When $\calD=\Hilb$, the functor defined in (\ref{eq:L-bulk-to-defect}) is just the bulk-to-boundary functor $Z(\calC) \to \calC_\calM^\vee$. We have already discussed this functor for the ``smooth'' and ``rough'' boundaries of the toric code model, see~(\ref{eq:bulk-sbound}) and~(\ref{eq:bulk-rbound}). For the transparent wall in Figure~\ref{toric-defect}, we have $\calC=\calD=\text{Rep}_{\ZZ_2}$,\, $\calM=\Hilb$, and the functors $L_\calM,R_\calM$ are given by~(\ref{eq:bulk-to-defect-toric}). When $\calC$ and $\calD$ are Morita equivalent via $\calM$, both functors $L_\calM$ and $R_\calM$ are monoidal equivalences, and the functor $T_\calM:=R_\calM^{-1}\circ L_\calM$ is an equivalence between braided tensor categories $Z(\calC)$ and $Z(\calD)$ \cite{eo}. It describes the tunneling of bulk excitations between the $\calC$ and $\calD$ regions (see also~(\ref{pic:transparent-wall})).

We see that a transparent $\calM$-wall naturally gives a braided equivalence $T_\calM: Z(\calC) \to Z(\calD)$ of two bulk phases. A much stronger statement is true: the correspondence $\calM \mapsto T_\calM$ between transparent domain walls and equivalences of bulk phases is one-to-one. More exactly, the class of all UTCs can be made into a groupoid (a category with invertible morphisms) in two ways: (i)~the morphisms between $\calC$ and $\calD$ are equivalence classes of bimodules, or (ii)~the morphisms are braided equivalences between $Z(\calC)$ and $Z(\calD)$. These two groupoids are denoted by $\mathcal{P}\text{ic}$ and $\mathcal{A}\text{ut}$, and the assignment $\calM \mapsto  T_\calM$ defines an isomorphism\footnote{The second author (LK) announced this result in his talk at the conference on topological field theory held in Northwestern University in May 2009. This is not an isolated phenomenon. A similar result, announced in the same talk, holds in 2-dimensional rational conformal field theory \cite{dkr}.} (obtained independently in \cite{enom09}): 
\be \label{eq:duality-defect}
T: \mathcal{P}\text{ic} \xrightarrow{\mbox{   $\cong$  } } \mathcal{A}\text{ut}. 
\ee
When restricted to a single phase associated with $\calC$, we obtain an isomorphism between the Picard group $\mathrm{Pic}(\calC)$ of all invertible $\calC$-$\calC$-bimodules and the group $\mathrm{Aut}(Z(\calC))$ of auto-equivalences of $Z(\calC)$.

For example, the group $\text{Pic}(\text{Rep}_{\mathbb{Z}_2})$ consists of two elements (invertible $\text{Rep}_{\mathbb{Z}_2}$-$\text{Rep}_{\mathbb{Z}_2}$-bimodules): $\text{Rep}_{\mathbb{Z}_2}$ and $\Hilb$. They correspond to the trivial domain wall and the wall depicted in Figure~\ref{toric-defect}. The inverse of $\Hilb$ is just itself: indeed, the $\Hilb$-wall simply interchanges the lattice and its dual. In tensor-categorical terms, $\Hilb\boxtimes_{\text{Rep}_{\mathbb{Z}_2}} \Hilb\cong \text{Rep}_{\mathbb{Z}_2}$ as $\text{Rep}_{\mathbb{Z}_2}$-$\text{Rep}_{\mathbb{Z}_2}$-bimodules.  The tunneling isomorphism through the $\Hilb$-wall is the braided equivalence $T_{\Hilb}: Z(\text{Rep}_{\mathbb{Z}_2})\rightarrow Z(\text{Rep}_{\mathbb{Z}_2})$ given by~(\ref{eq:toric-tunnel}).  Thus, $\mathrm{Aut}(Z(\text{Rep}_{\mathbb{Z}_2}))\cong \text{Pic}(\text{Rep}_{\mathbb{Z}_2}) \cong \mathbb{Z}_2$.

If a ${}_\calC\calM_\calD$-wall is not transparent, both functors $L_\calM$ and $R_\calM$ are not invertible. In this case, excitations cannot generally cross the wall. However, we can imagine a process where a droplet of $\calC$ phase containing an excitation (depicted by $\times$) is injected into the $\calD$ phase:
\[
 \raisebox{-50pt}{
  \begin{picture}(100,110)
   \put(50,8){\scalebox{1.5}{\includegraphics{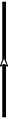}}}
   \put(50,8){
     \setlength{\unitlength}{.75pt}\put(-151,-235){
     \put(110,330)  {\scriptsize $ \calC $}
     \put(185,330)  {\scriptsize $ \calD $}
     \put(148,355)  {\scriptsize $ \calM $}
     \put(148,225)  {\scriptsize $\calM $}
     \put(110, 290) {\scriptsize $\times$}
     }\setlength{\unitlength}{1pt}}
  \end{picture}}
  \rightsquigarrow
   \raisebox{-50pt}{
   \begin{picture}(130,110)
   \put(50,8){\scalebox{1.5}{\includegraphics{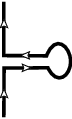}}}
   \put(50,8){
     \setlength{\unitlength}{.75pt}\put(-151,-235){
     \put(120,290)  {\scriptsize $ \calC $}
     \put(170,310)  {\scriptsize $ \calD $}
     \put(170,270)  {\scriptsize $ \calD $}
     \put(148,355)  {\scriptsize $ \calM $}
     \put(148,225)  {\scriptsize $\calM $}
     \put(203, 290) {\scriptsize $\times$}
     }\setlength{\unitlength}{1pt}}
  \end{picture}}
  \rightsquigarrow
  \raisebox{-50pt}{
   \begin{picture}(130,110)
   \put(50,8){\scalebox{1.5}{\includegraphics{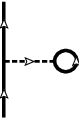}}}
   \put(50,8){
     \setlength{\unitlength}{.75pt}\put(-151,-235){
     \put(120,290)  {\scriptsize $ \calC $}
     \put(170,330)  {\scriptsize $ \calD $}
     \put(170,250)  {\scriptsize $ \calD $}
     \put(173,299)  {\scriptsize $\oplus_i \calN_i$}
     \put(148,355)  {\scriptsize $ \calM $}
     \put(148,225)  {\scriptsize $\calM $}
     \put(230,288) {\scriptsize $\calM $}
     \put(210, 288) {\scriptsize $\times$}
     }\setlength{\unitlength}{1pt}}
  \end{picture}}
\]
The double wall in the middle picture is resolved as a sum of indecomposable $\calD$-$\calD$-bimodules:
\[
\calM^\op \boxtimes_\calC \calM = \calC_\calM^\vee= \oplus_{i=1}^N \calN_i.
\]
For example if $\calD=\Hilb$, then $\calC_\calM^\vee$ is a unitary category with trivial module structure (i.e.\ a bimodule over the trivial UTC $\Hilb$). In this case, it can be decomposed into several pieces $\calN_i$, namely, copies of $\Hilb$ associated with the simple objects $i\in\calC_\calM^\vee$. Thus, when a bulk excitation is forced through a non-invertible domain wall (by a suitable alteration of the model), it pulls a superposition of defect lines labeled by $\calN_i$. This is a categorified version of the tunneling through topological defects in rational conformal field theories \cite{ffrs}. 

However, if $\calM$ is invertible, then $\calC_\calM^\vee=\calD$ is regarded as a bimodule over itself, and the only $\calN_i$ appearing in the direct sum is $\calD$. Thus, we simply have the following tunneling process: 
\be \label{pic:transparent-wall}
 \raisebox{-50pt}{
  \begin{picture}(100,110)
   \put(50,8){\scalebox{1.5}{\includegraphics{C-M-D}}}
   \put(50,8){
     \setlength{\unitlength}{.75pt}\put(-151,-235){
     \put(110,330)  {\scriptsize $ \calC $}
     \put(185,330)  {\scriptsize $ \calD $}
     \put(148,355)  {\scriptsize $ \calM $}
     \put(148,225)  {\scriptsize $\calM $}
     \put(110, 290) {\scriptsize $\times$}
     }\setlength{\unitlength}{1pt}}
  \end{picture}}
  \rightsquigarrow
   \raisebox{-50pt}{
   \begin{picture}(130,110)
   \put(50,8){\scalebox{1.5}{\includegraphics{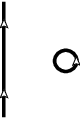}}}
   \put(50,8){
     \setlength{\unitlength}{.75pt}\put(-151,-235){
     \put(120,290)  {\scriptsize $ \calC $}
     \put(170,290)  {\scriptsize $ \calD $}
     \put(148,355)  {\scriptsize $ \calM $}
     \put(148,225)  {\scriptsize $\calM $}
     \put(230,288) {\scriptsize $\calM $}
     \put(209, 290) {\scriptsize $\times$}
     }\setlength{\unitlength}{1pt}}
  \end{picture}}
\ee
It is very easy to see from the picture that this tunneling is given by a monoidal functor because
$$
 \raisebox{-20pt}{
  \begin{picture}(140,50)
   \put(50,8){\scalebox{1.5}{\includegraphics{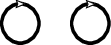}}}
   \put(50,8){
     \setlength{\unitlength}{.75pt}\put(-151,0){
     \put(165, 17) {$\times$}
     \put(233, 17)   {$+$}
     }\setlength{\unitlength}{1pt}}
  \end{picture}}
  =
  \raisebox{-20pt}{
  \begin{picture}(110,50)
   \put(10,10){\scalebox{1.5}{\includegraphics{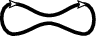}}}
   \put(10,10){
     \setlength{\unitlength}{.75pt}\put(-151,0){
     \put(165, 13) {$\times$}
     \put(225, 13)   {$+$}
     }\setlength{\unitlength}{1pt}}
  \end{picture}}
  =
   \raisebox{-20pt}{
  \begin{picture}(80,50)
   \put(10,8){\scalebox{1.5}{\includegraphics{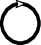}}}
   \put(10,8){
     \setlength{\unitlength}{.75pt}\put(-151,0){
     \put(159, 17) {$\times$}
     \put(171, 17)   {$+$}
     }\setlength{\unitlength}{1pt}}
  \end{picture}}
$$ 
where $\times$ and $+$ are two bulk excitations. Moreover, the tunneling process preserves the double braiding of anyonic excitations. This operation of encircling excitations by an invertible defect line $\calM$ is a physical realization of the tunneling isomorphism $T_\calM$, which is an equivalence of braided tensor categories between $Z(\calC)$ and $Z(\calD)$. Its inverse can be obtained by further encircling the resulting object by an $\calM^\op$ line. This gives an informal explanation of the fact that $Z(\calC) \cong Z(\calD)$ as braided tensor categories if $\calC$ and $\calD$ are Morita equivalent. Conversely, it was proved in \cite{eno08} that $\calC$ and $\calD$ are Morita equivalent if $Z(\calC) \cong Z(\calD)$ as braided tensor categories.

\section{Defects of higher codimensions} \label{sec:defect-cod-23}

In Section \ref{sec:lw-defect}, we considered excitations on a single  ${}_\calC\calM_\calD$ domain wall. More generally, different types of wall between the $\calC$ and $\calD$ bulk phases may join at a point. We call the boundary between a ${}_\calC\calM_\calD$-wall and a ${}_\calC\calN_\calD$-wall an \emph{$\calM$-$\calN$-excitation}. Such excitations correspond to representations of an operator algebra denoted by $A_{\calM,\calN}$:
\[
A_{\calM,\calN} := \bigoplus_{j \in \calC\boxtimes\calD^\opt}\, 
\bigoplus_{\lambda,\sigma\in \calM}\, \bigoplus_{\gamma,\rho\in \calN} 
\Hom_\calM(j\otimes \sigma, \lambda) \otimes \Hom_\calN(\gamma, j\otimes \rho)\qquad \left(\text{graphically,}\quad\figbox{1.0}{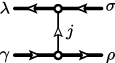}\right),
\]
where $j\in\calC\boxtimes\calD^\opt$,\,\, $\lambda,\sigma\in\calM$, and $\gamma,\rho\in\calN$ run over the simple objects.  The multiplication and the unit in $A_{\calM,\calN}$ are defined like those in $A$, see Section \ref{sec:lw-boundary}. There is no bialgebra structure on $A_{\calM,\calN}$ (though one can define a map $A_{\calM,\calN}\to A_{\calM,\calK}\otimes A_{\calK,\calN}$ for any $\calK$). As a consequence, the category of $A_{\calM, \calN}$-modules is not monoidal. This category is equivalent to $\Fun_{\calC|\calD}(\calM, \calN)$, the category of $\calC$-$\calD$-bimodule functors from $\calM$ to $\calN$.

Although two excitations in $\Fun_{\calC|\calD}(\calM,\calN)$ can not fuse (just because they can not be placed next to each other), there is a well-defined fusion of an $\calM$-$\calK$-excitation with a $\calK$-$\calN$-excitation that produces an $\calM$-$\calN$-excitation. It is given by the composition of functors (extending the definition of the tensor product in $\calC_\calM^\vee$): 
\begin{equation}\label{eq:wallex_fusion}
\otimes:\,\Fun_{\calC|\calD}(\calM,\calK)\times \Fun_{\calC|\calD}(\calK,\calN)
\to \Fun_{\calC|\calD}(\calM,\calN).
\end{equation}
Restricting to special cases, it is easy to see that this product makes the category $\Fun_{\calC|\calD}(\calM, \calN)$ into a $\Fun_{\calC|\calD}(\calM, \calM)$-$\Fun_{\calC|\calD}(\calN, \calN)$-bimodule. For each object $\calF\in\Fun_{\calC|\calD}(\calM,\calN)$ we have two functors, $\calF^*:X\mapsto X\otimes F$ and $\calF_*:X\mapsto F\otimes X$,
\[
\Fun_{\calC|\calD}(\calM, \calM) \xrightarrow{\calF^*} \Fun_{\calC|\calD}(\calM, \calN)
\xleftarrow{\calF_*}  \Fun_{\calC|\calD}(\calN, \calN).
\]
Both $\calF^*$ and $\calF_*$ are $Z(\calC)$-$Z(\calD)$-bimodule functors, and the following diagram is commutative up to invertible natural transformations:
\[
\xymatrix{
 & \Fun_{\calC|\calD}(\calM, \calM) \ar[d]^{\calF^*} & \\
Z(\calC) \ar[ru]^{L_\calM\,} \ar[rd]_{L_\calN}
& \Fun_{\calC|\calD}(\calM, \calN) &
Z(\calD) \ar[lu]_{R_\calM} \ar[ld]^{\,R_\calN} \\
 & \Fun_{\calC|\calD}(\calN, \calN) \ar[u]_{\calF_*} & 
}
\]

The above structures can be seen explicitly in the toric code model. Recall that the endpoint of the defect line in Figure~\ref{toric-defect} has an associated operator $Q$ defined by~(\ref{eq:Q}). Its eigenvalues correspond to two distinct simple objects in the category of bimodule functors, $F_{+},F_{-}\in\Fun_{\Rep_{\ZZ_2}|\Rep_{\ZZ_2}}(\Rep_{\ZZ_2},\Hilb)$. The other end of the defect line is described by the adjoint functors $\dual{F}_{+}$, $\dual{F}_{-}$, and the fusion (or composition of functors according to~(\ref{eq:wallex_fusion})) is as follows:
\begin{equation}
F_{+}\otimes\dual{F}_{+}=F_{-}\otimes\dual{F}_{-}
\cong 1 \oplus \epsilon,
\qquad\quad
F_{+}\otimes\dual{F}_{-}=F_{-}\otimes\dual{F}_{+}
\cong e \oplus m.
\end{equation}

All types of excitations, including those connecting two different walls, can be viewed as defects of codimension 2. The most general defect of codimension 2 is a junction of multiple domain walls. For example, a 3-way junction is defined by UTCs $\calC_1$, $\calC_2$, $\calC_3$ and unitary categories $\calM_{12}$ (a $\calC_1$-$\calC_2$-bimodule), $\calM_{23}$ (a $\calC_2$-$\calC_3$-bimodule), and $\calM_{13}$ (a $\calC_1$-$\calC_3$-bimodule). The junction is characterized by an arbitrary object 
\begin{equation}\label{eq:3-way}
\calF \in \Fun_{\calC_1|\calC_3}\bigl(\calM_{13},\,
\calM_{12} \boxtimes_{\calC_2} \calM_{23}\bigr).\qquad\qquad
\figbox{1.0}{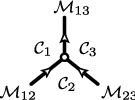}
\end{equation}

Can we have defects of even higher codimension? Yes indeed, the physical system also evolves in time and should be described by a $2+1$ theory. This time evolution provides the last layer of structure. A codimension 3 defect is an event that occurs at some point in space-time and is described by a natural transformation $\phi: \calF \to \calG$ between two bimodule functors. For example, one can imagine a process in which an excitation $F$ on the $\calC_1$-$\calC_2$ domain wall (a special case of codimension 2 defect) annihilates with its anti-excitation $\overline{F}$ (the adjoint functor, or monoidal dual object). A partial description of the initial and final quantum states is given by the objects $F\otimes\overline{F}$ and $\one$, respectively. In our formalism, the annihilation is characterized by a natural transformation $\phi: F\otimes \overline{F}\to\one$, which is represented by a certain operator acting in the Hilbert space of the whole system. The fusion of codimension 3 defects is given by the composition of natural transformations; physically, it corresponds to one event occuring after another.

\section{Summary and outlook}  \label{sec:dictionary}

We have extended the Levin-Wen model by adding external boundaries and defects of codimension 1, 2, 3, while utilizing the entire representation theory of unitary tensor categories. The correspondence between physical notions and the tensor-categorical formalism is summarized in Table~\ref{tab:dict}.

\begin{table}[t]
\centerline{\begin{tabular}{| p{7.3cm} | p{8.2cm} |}
      \hline
      Ingredients of Levin-Wen models  & Tensor-categorical notions \\  \hhline{|=|=|}
      bulk Levin-Wen model  &  unitary tensor category $\calC$  \\ \hline
      edge labels in the bulk & simple objects in $\calC$  \\  \hline
      excitations in the bulk & objects in $Z(\calC)$, the monoidal center of $\calC$ \\ \hhline{|=|=|}
      boundary type & $\calC$-module $\calM$ \\ \hline
      edge labels on a ${}_\calC\calM$-boundary & simple objects in $\calM$ \\ \hline
      excitations on a ${}_\calC\calM$-boundary & objects in the category $\mathrm{Fun}_\calC(\calM, \calM)$ of $\calC$-module functors \\ \hline 
      bulk excitations fusing into\newline a ${}_\calC\calM$-boundary  & $Z(\calC)=\Fun_{\calC|\calC}(\calC, \calC) \rightarrow \Fun_\calC(\calM, \calM)$\newline $(\calC \xrightarrow{\calF} \calC) \mapsto (\calC \boxtimes_\calC \calM \xrightarrow{\calF \boxtimes \id_\calM} \calC \boxtimes_\calC \calM)$. \\ \hhline{|=|=|}
      domain wall & $\calC$-$\calD$-bimodule $\calN$  \\ \hline
      edge labels on a ${}_\calC\calN_\calD$-wall  & simple objects in $\calN$ \\ \hline
      excitations on a ${}_\calC\calN_\calD$-wall & objects in the category $\mathrm{Fun}_{\calC|\calD}(\calN, \calN)$ of $\calC$-$\calD$-bimodule functors \\ \hline
      fusion of two walls & $\calM \boxtimes_\calD \calN$  \\  \hline
      invertible ${}_\calC\calN_\calD$-wall  & $\calC$ and $\calD$ are Morita equivalent, i.e.\newline $\calN\boxtimes_\calD \calN^\op \cong \calC$,\,\, $\calN^\op \boxtimes_\calC \calN \cong \calD$.  \\ \hhline{|=|=|}
      defects of codimension 2 ($\calM$-$\calN$-excitations)
      & objects $\calF, \calG \in \Fun_{\calC|\calD}(\calM, \calN)$ \\ \hhline{|=|=|}
      defect of codimension 3 
      & natural transformation $\phi: \calF \to \calG$ \\  \hline
   \end{tabular}}
\caption{Dictionary between ingredients of Levin-Wen models and tensor-categorical notions.}
\label{tab:dict}
\end{table}

Our constructions are related to the so-called extended Turaev-Viro topological field theory \cite{TV}\cite{bw}\cite{kirillov-balsam}\cite{lurie}\cite{kapustin}, in which the unitary tensor category $\calC$ (or even better, the bicategory of $\calC$-modules) is assigned to a point, a bimodule is assigned to a framed interval, the category $Z(\calC)$ is assigned to a circle, etc. The last step is to assign a number, called the Turaev-Viro invariant to a 3-manifold (possibly, with corners). We envision a more general scheme, though we do not make any claim regarding its topological invariance.

Let us consider a 3-cell complex $\Lambda$ (a space-time) in which the map from the boundary of each $d$-cell to the $d-1$ skeleton is locally one-to-one. We
inductively associate algebraic objects with the cells of $\Lambda$, descending from dimension 3 to 0:
\begin{itemize}
\item[3)] Each 3-cell $i$ is assigned an arbitrary UTC $\calC_i$.
\item[2)] A 2-cell $j$ that has 3-cells $i_1,\dots,i_r$ attached to it is assigned a module $\calM$ over $\calC_{i_1}\boxtimes\dots\boxtimes\calC_{i_r}$. (Depending on the orientation, some of the $\calC$'s should be replaced by the opposite categories.)
\item[1)] A 1-cell $k$ at the junction of 2-cells $j_1,\dots,i_s$ is assigned an object $\calF$ in the unitary category that generalizes the construction~(\ref{eq:3-way}). 
\item[0)] A 0-cell $l$ is assigned a vector in some Hilbert space that depends on the incident 1-, 2-, and 3-cells. This step is nontrivial, and we have not worked it out in detail. One needs to look at a small neighborhood of $l$, which is a cone over a 2-dimesional cell complex. The Hilbert space is defined by that complex and the associated algebraic data. In the simplest case where the complex is a sphere made of 2-cells $\calC,\calD$, 1-cells $\calM,\calN$, and 0-cells $\calF,\calG\in\Fun_{\calC|\calD}(\calM,\calN)$,  the space in question consists of natural transformations between $\calF$ and $\calG$.
\end{itemize}
We further conjecture that whole tower defines a complex number (a kind of partition function).

It would also be interesting to see how Levin-Wen models can be generalized to higher dimensions. In particular, a generalization of the Turaev-Viro theory to 3+1 dimensions is known \cite{ma}. In this case, one uses data in a spherical 2-category as the building blocks. For even higher dimensional theories, Lurie's classification of extended TQFTs \cite{lurie} suggests that one can start with a so-called fully dualizable object in a higher category to build the theory.

\end{document}